\documentclass[twoside]{LCWS11}
\usepackage[latin1]{inputenc}
\usepackage{graphicx}
\usepackage{wrapfig,rotating}
\usepackage{amssymb,amsmath,array}
\usepackage{cite}
\begin{document}
\title{
%%%%   Paper title goes here  %%%%%%%%%%%%%%
Shower Leakage Correction in a High Granularity Calorimeter} %% 
%***********************************************************************
% AUTHORS INFORMATION AREA
%***********************************************************************
\author{Shaojun Lu, on behalf of CALICE collaboration
% Optional short acknowledgment: remove next line if non-needed
%\thanks{This is an optional funding source acknowledgment.}
% DO NOT MODIFY THE FOLLOWING '\vspace' ARGUMENT
\vspace{.3cm}\\
% Addresses and institutions (remove "1- " in case of a single institution)
Deutsches Elektronen-Synchrotron (DESY) \\
Notekstrasse 88, 22607, Hamburg, Germany
%% Remove the next three lines in case of a single institution
}
%%***********************************************************************
% END OF AUTHORS INFORMATION AREA
%***********************************************************************

\maketitle

\begin{abstract}
In the ILD detector,  HCAL is inside the coil. The HCAL is about 5 interaction length thick, and ECAL is about 1 interaction length thick.
For example, an eighty GeV hadron, it is only 95\% energy will be contained in ECAL plus HCAL.
We need a topological reconstruction of the leakage.
This has been studied with experimental data collected using the CALICE prototypes, during the 2007 test beam campaign at CERN. The complete setup of the experiment consisted of a silicon-tungsten electromagnetic calorimeter, an analog scintillator-steel hadron calorimeter and a scintillator-steel tail catcher. Events collected using pion beams in the energy range 8-100~GeV are selected and compared to the Monte Carlo simulations. While the leakage from the full setup is negligible, when removing the tail catcher information either partly or completely the energy loss becomes significant and affects the performance. The average measured energy decreases below the expected beam energy and the resolution deteriorates. A correction to the leakage was implemented for pions. The results obtained show that the correction is powerful in restoring the mean value of the measured energy distributions back to the expected beam energy, with an accuracy at the 1-2\% level over the whole energy range. The relative improvement on the resolution is about 25\% at 80~GeV, decreasing at lower energies together with the impact of the leakage.
\end{abstract}

\section{The AHCAL prototype detector}

\begin{wrapfigure}{r}{0.5\columnwidth}
\centerline{\includegraphics[scale=0.6]{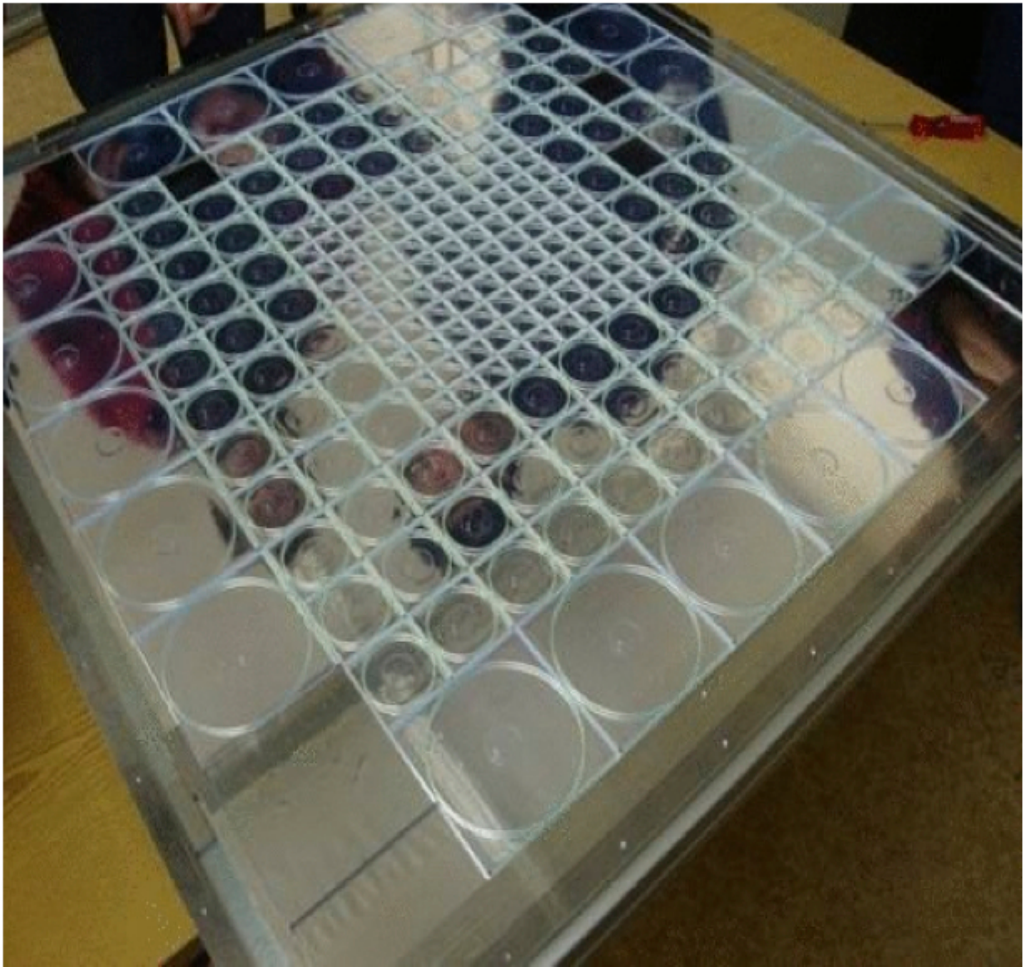}}
\caption{Picture of one AHCAL active layer}\label{Fig:MV}
\end{wrapfigure}

The goal of the CALICE experimental program is to establish novel technologies for calorimetry in future collider experiments and to record electromagnetic and hadronic shower data with unprecedented three dimensional spatial resolution for the validation of simulation codes and for the test and development of reconstruction algorithms. Such highly granular calorimeters are necessary to achieve an unprecedented jet energy resolution at the International Linear Collider~\cite{ILC} using particle flow algorithms.~\cite{PFA} The CALICE analog hadron calorimeter (AHCAL)~\cite{AHCAL} prototype is a 38 layers sampling calorimeter, which is built out of scintillator tiles with sizes ranging from $30\times30~\mathrm{mm}^2$ in the core of the detector to $120\times120~\mathrm{mm}^2$ . The light in each scintillator cell is collected by a wavelength shifting fiber, which is coupled to a silicon photomultiplier (SiPM) SiPM. The SiPMs are produced by the MEPhI/PULSAR group.~\cite{SiPM} They have a photo-sensitive area of $1.1\times1.1~\mathrm{mm}^2$
containing 1156 pixels with a size of $32\times32~\mathrm{\mu m}^2$. In total, the calorimeter has 7608 channels. A built-in LED calibration system with UV LEDs, is coupled to each cell by clear fibers, and equipped with PIN diodes to monitor the LED light intensity. The performance of the AHCAL was validated with positrons at various energies~\cite{AHCAL}. Good linearity and satisfactory agreement with simulation up to an energy of 50~GeV has been observed.~\cite{ELECTRON,CAN014}

%An analog hadron calorimeter (AHCAL) \cite{AHCAL} prototype of 5.3 nuclear interaction lengths thickness has been designed and constructed by members of the CALICE Collaboration. The AHCAL prototype consists of a 38-layer sandwich structure of steel plates and 7608 scintillator tiles that are read out by wavelength-shifting fibres coupled to SiPMs. The signal is amplified and shaped with a custom-designed ASIC. A calibration/monitoring system based on LED light was developed to monitor the SiPM gain and to measure the full SiPM response curve in order to correct for non-linearity. Ultimately, the physics goals are the study of hadronic shower shapes and testing the concept of particle flow. The technical goal consists of measuring the performance and reliability of 7608 SiPMs. The AHCAL prototype was commissioned in test beams at DESY, CERN and FNAL, and recorded hadronic showers, electron showers and muons at different energies and incident angles.

\section{Identification of track segments}
One fine granular AHCAL prototype layer is shown in figure~\ref{Fig:MV}. 
The high granularity of the active layers in the hadronic calorimeter and the cell-by-cell readout gives the CALICE detectors unprecedented imaging capabilities. 

\begin{wrapfigure}{r}{0.5\columnwidth}
%\begin{figure}
\centerline{\includegraphics[scale=0.4]{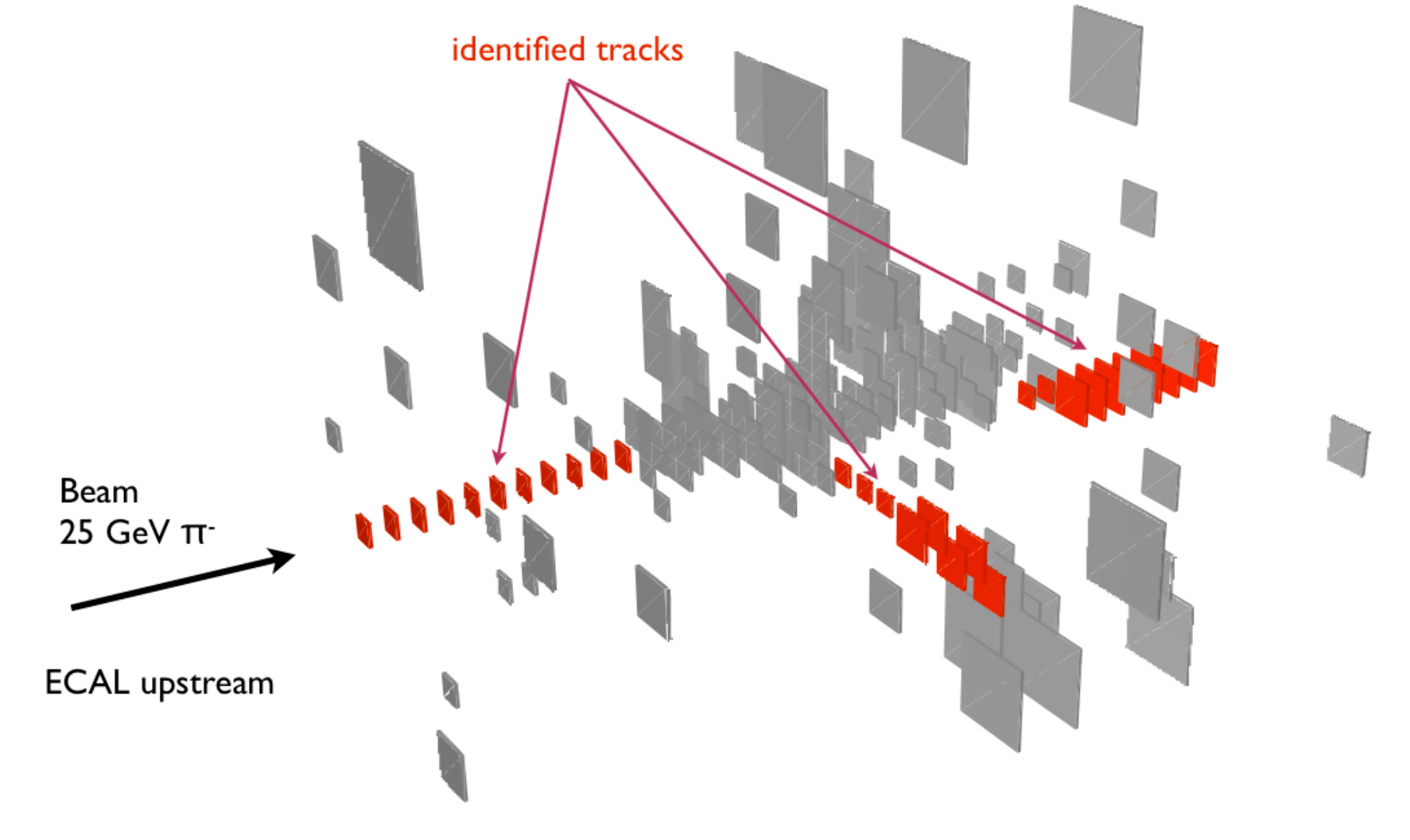}}
\caption{Identification of minimum-ionizing track segments within hadronic showers.}\label{Fig:Seg}
%\end{figure}
\end{wrapfigure}

Thanks to the high granularity, the development of hadronic showers can be studied in great detail. The response of the AHCAL to hadronic showers are investigated.
This is exploited to study the topology of hadronic events in detail. Track segments created by charged particles produced within the hadronic shower can be identified, provided the particles travel an appreciable distance before interacting again and are separated from other activity in the detector.

Figure~\ref{Fig:Seg} demonstrated that three tracks were found within one 25~GeV $\pi$ event. The properties of these tracks are sensitive to the substructure of the hadronic shower, and can thus serve as a powerful probe for hadronic shower models. The track segments identified in hadronic showers have a high quality, and are also suitable for detector calibrations via the extraction
of the most probable value of the energy loss in each cell along the track.~\cite{TRACKSEGMENT}

% In the CALICE detector prototype, this technique was used to study the temperature dependence of the detector response.

\section{Longitudinal shower profile}

\begin{figure}
\centerline{\includegraphics[scale=0.5]{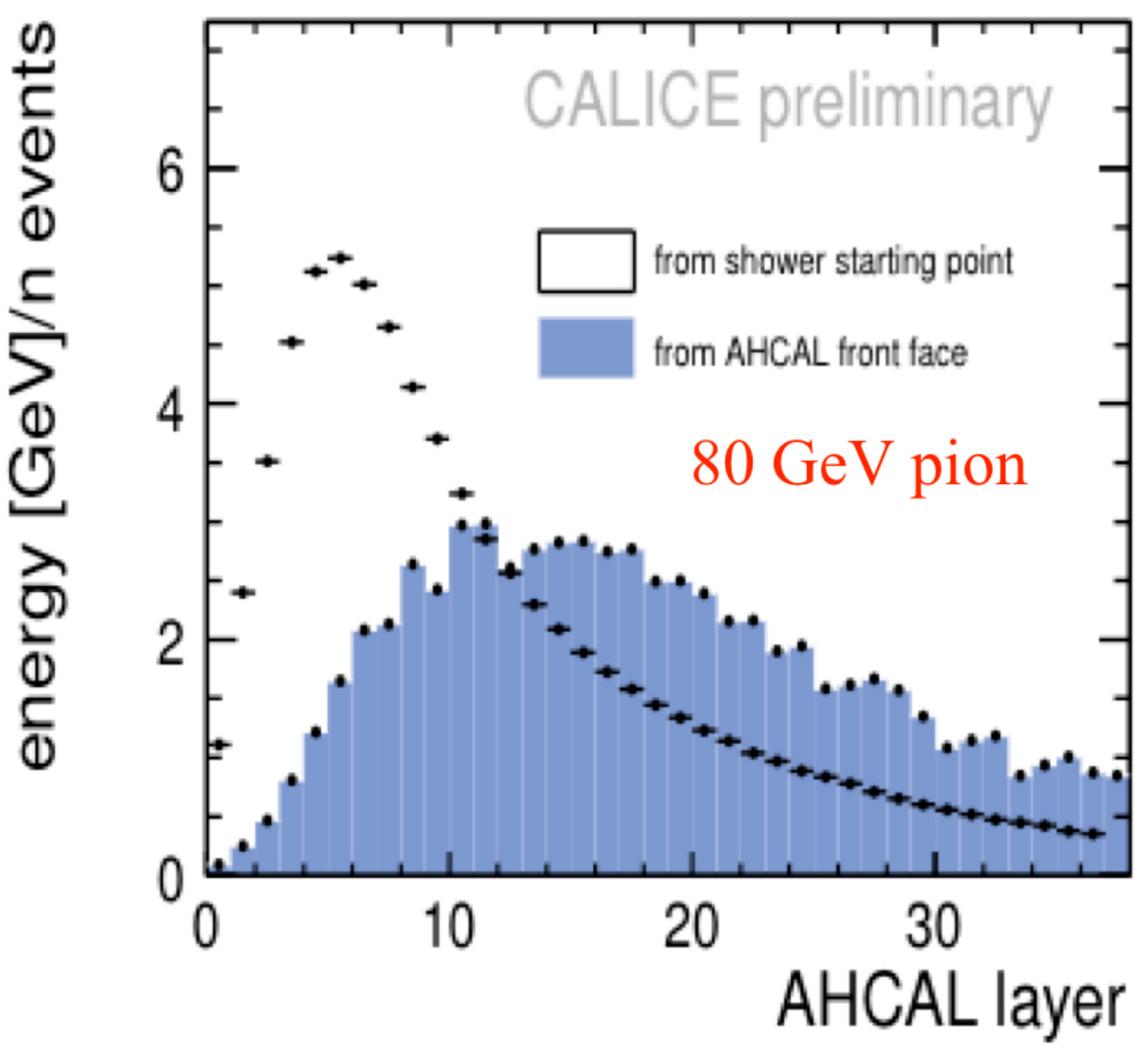}}
\caption{longitudinal profile in the AHCAL only, drawn with respect to the first hard interaction of the showers or w.r.t. the calorimeter front face. Showers starting in the first layer of the AHCAL have been rejected, therefore the longitudinal profile relative to the shower starting point does not reach the very last bin.}\label{Fig:ShowerStart}
\end{figure}

Hadronic shower development is based largely on nuclear interactions, and therefore hadrons can travel a significant distance before a shower develops. The average distance a hadron travels before a nuclear interaction occurs, which typically starts the cascade of the shower, is called nuclear interaction length $\lambda_{int}$.

The high granularity of the calorimeter provide the possibility to determine the first hard interaction of the showers, it is possible to reconstruct the development of the showers relative to their starting point, instead of considering the development relative to the calorimeter front face. When the shower starting point for each event is known, it is possible to shift the single event profile to the same point. The plot in Figure~\ref{Fig:ShowerStart} shows the longitudinal profile for 80~GeV pion showers. These two ways to reconstruct the longitudinal profile are compared.  The open symbols indicate that the longitudinal shower profile becomes much shorter after correcting event-by-event for the variation of the shower starting point. The layer-by-layer effects are clearly washed out and the profile development appears to be smooth.

%From AHCAL prototype test beam data, the shower starting point as a function of the number of interaction length is studied, the result is shown in the left plot of figure ?. The decreasing exponential slope agrees with expectations. 

The measurement of the position of the first hadron interaction allows to disentangle the fluctuations of the position where the shower statrts from those intrinsic in the shower process.

\section{Leakage correction}

\subsection{The leakage}

\begin{figure}
\centerline{\includegraphics[scale=0.6]{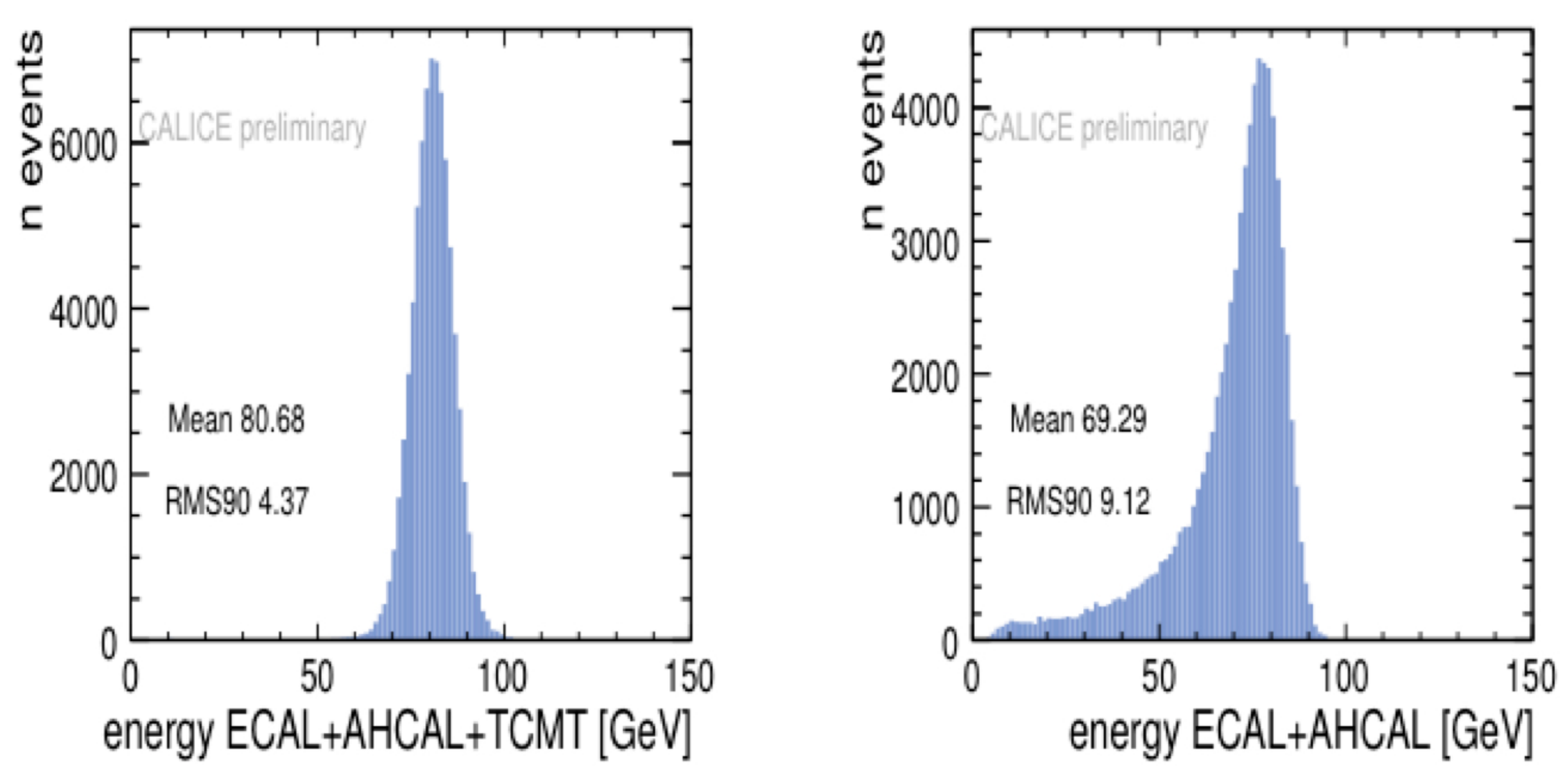}}
\caption{Left: total energy measured using the full calorimeter. Right: energy measured by SiW-ECAL+AHCAL only.}\label{Fig:ShowerLeakage}
\end{figure}

In figure~\ref{Fig:ShowerLeakage}, the distribution on the left plot shows the total energy measured using the full calorimeter, while the distribution on the right plot shows the energy measured when excluding the TCMT information. To quantify the energy resolution, the measure RMS90 has been used, defined in~\cite{PFA} as the root mean square of the smallest interval containing 90\% of the the distribution. The degradation is clearly visible on the distribution when excluding the TCMT information. It shows both in terms of mean and RMS90 in the distribution on the right plot. A topological reconstruction of the leakage has been studied and implemented to correct the leakage.

\subsection{Observables sensitive to the leakage}

\begin{figure}
\centerline{\includegraphics[scale=0.5]{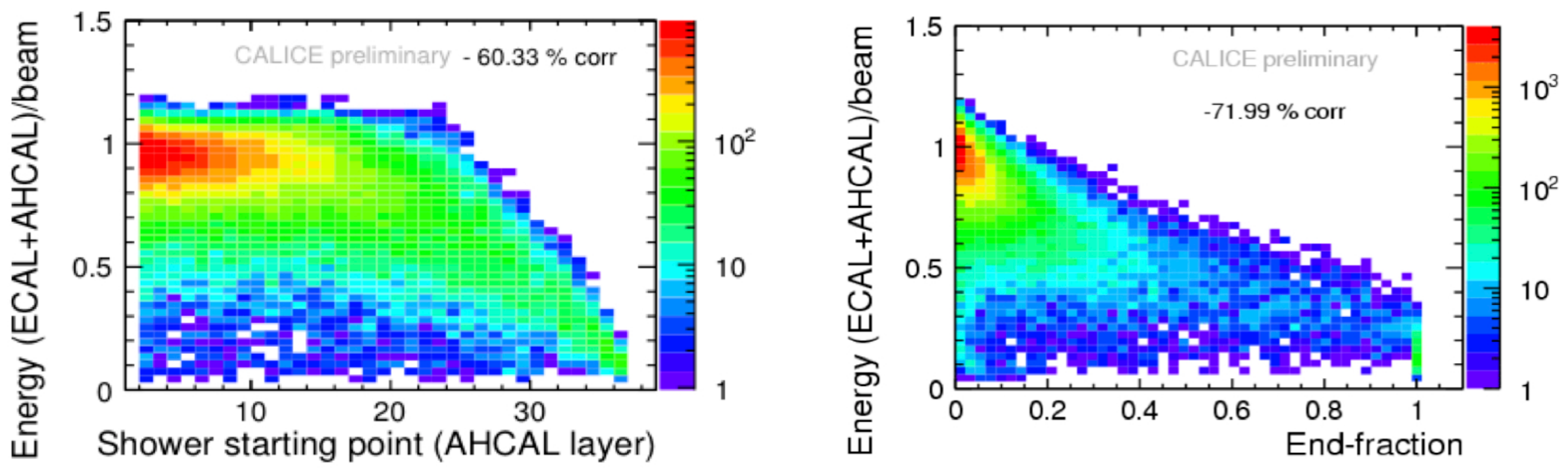}}
\caption{Left: shower starting point versus the ratio between the measured energy in SiW- ECAL+AHCAL and the beam energy. Right: end-fraction versus the ratio between the energy measured in SiW- ECAL+AHCAL and the beam energy. The cluster of events for an end-fraction of approximately one is due to the fact that events with a shower starting point in layers 35 or 36 have an end-fraction exactly equal to one according to the end-fraction definition.}\label{Fig:LookingupTable}
\end{figure}

Since the information of the TCMT is excluded, measured energy will indicate the energy measured by SiW-ECAL+AHCAL, where not otherwise specified. Two observables have been studied which are sensitive to the leakage. They are shown in figure~\ref{Fig:LookingupTable}.  The left distribution shows the correlation between the shower starting point and the ratio between the measured energy and the beam energy. When the shower starts at the beginning of the AHCAL, the measured energy is close to the expected beam energy for most of the events. The leaking energy from the AHCAL, is usually negligible. On the contrary, for showers starting late in the AHCAL, the fraction of the beam energy which is measured becomes smaller. That means only partial energy deposited in the AHCAL. And the effect of the leakage is non-negligible anymore. As already mentioned, due to event-to-event fluctuations typical for hadronic showers, the correlation between the leakage and the shower starting point is spoiled, resulting in a broad distribution.

In order to recover the incident particle energy from the leakage, the distribution on the left in figure~\ref{Fig:LookingupTable} shows the average correction to be applied to the measured energy, depending on the layer where the first hard interaction occurred.

The right distribution in figure~\ref{Fig:LookingupTable} shows the end-fraction versus the ratio between the measured energy and the beam energy, for an 80~GeV run. The end-fraction is defined as the fraction of energy deposited in the last four layers of the AHCAL divided by the total shower energy measured by SiW-ECAL+AHCAL. A small fraction of energy deposited at the end of the AHCAL indicates that the shower is in general almost concluded before reaching the end of the calorimeter. In fact, the ratio between the measured energy and the beam energy is mainly close to one, for those events that have a small end-fraction. On the contrary, a large end-fraction indicates a shower still in development, that could leak into the TCMT with a high probability. As a consequence, the ratio between the measured and the beam energy decreases when the end-fraction increases, due to increasing leakage.

\subsection{A correction for the Leakage}
A correction method has been investigated and implemented within the CALICE collaboration by Ivan Marchesini for the leakage.~\cite{CAN029}
The correlation between the leakage and the shower starting point, the end-fraction has been used together in order to develop a realistic correction for the leakage. The shower starting point is calculated as the layer. The end-fraction is a continuous variable and can be arbitrarily binned. The chosen binning is [0, 0.01, 0.025, 0.05, 0.1, 0.15, 0.2, 0.25, 0.3, 0.35, 0.4, 0.45, 0.5, 0.55, 0.6, 0.65, 0.7, 0.75, 0.8, 0.85, 0.9, 0.95, 1]. 
The sum of measured energy in SiW-ECAL and AHCAL has been used to replace the beam energy information. In this case realistic means that the correction should not be tuned using the beam energy information.
The energy can be corrected event-by-event depending on the measured energy, the shower staring point and the end-fraction.

\subsection{Correction performance}

The correction was first implemented in a Monte Carlo study. Two possible applications to the data have then been considered, obtaining the correction to the data either from a Monte Carlo template or from independent data runs.

\begin{figure}
\centerline{\includegraphics[scale=0.6]{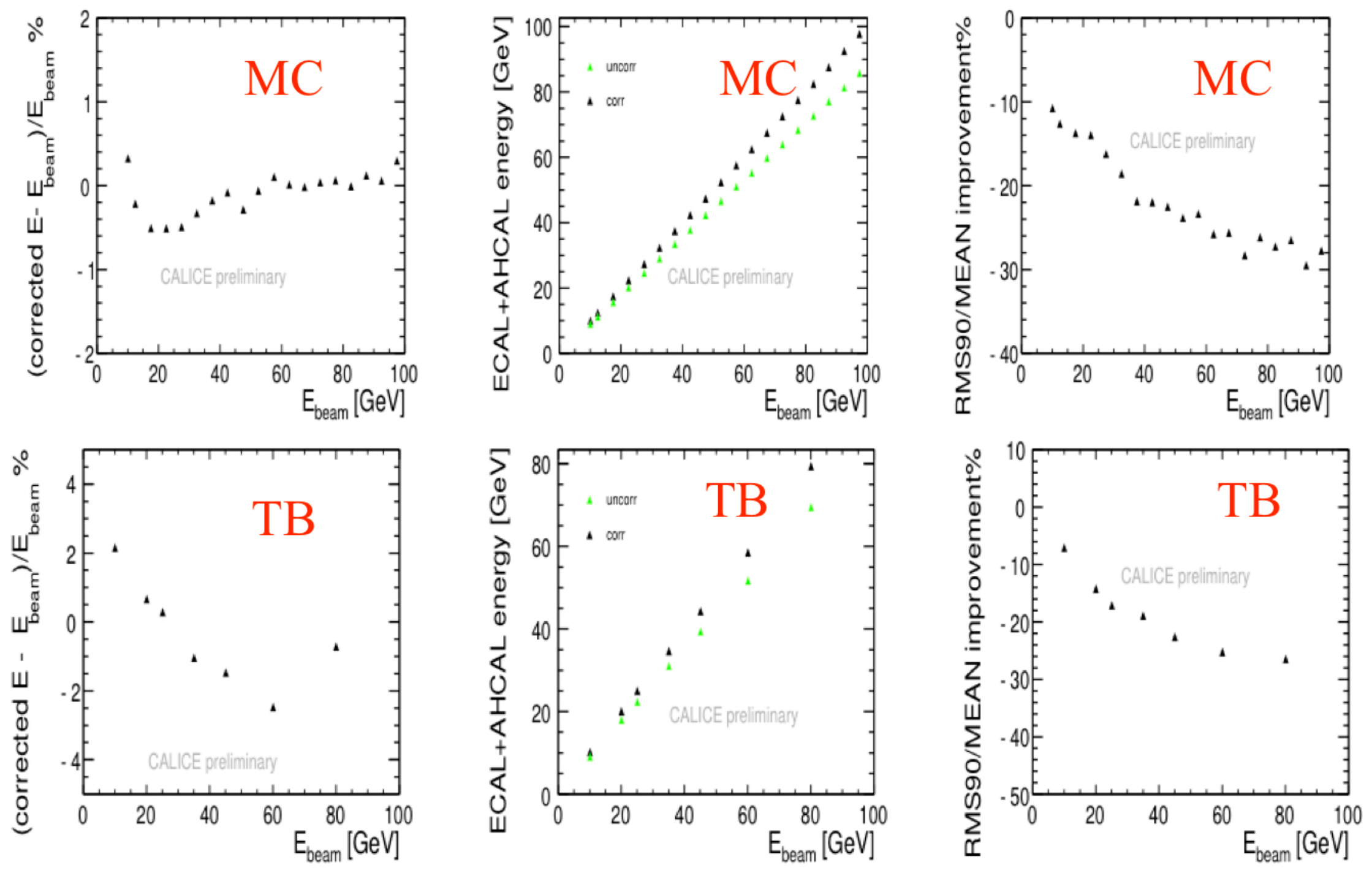}}
\caption{Top: Final results of the leakage correction for Monte Carlo. Left: the relative deviation of the mean value of the corrected energy distributions from the beam energy versus the beam energy. Center: mean value of the measured energy distributions before and after the correction. Right: relative improvement of RMS90/Mean versus the beam energy. Bottom: Final results of the leakage correction applied to data runs. Left: the relative deviation of the mean value of the corrected energy distributions from the beam energy versus the beam energy. Center: mean value of the measured energy distributions before and after the correction. Right: relative improvement of RMS90/Mean versus the beam energy. The templates for the correction have been obtained from independent data runs.}\label{Fig:LeakageCorrection}
\end{figure}

The performance of the correction is summarized on the top row in figure~\ref{Fig:LeakageCorrection} for the Mont Carlo studies. The correction is capable of identifying the energy of the run with a precision better than 0.5\%. The RMS90/Mean improves by up to 30\% at high energies, while decreasing towards lower energies together with the impact of the leakage.

The performance of the correction on the test beam data is shown on the bottom row in figure~\ref{Fig:LeakageCorrection}.
Pion data collected using the CALICE prototypes during the 2007 test beam campaign at CERN were analyzed.
When applying the correction to data, the looking up table can be obtained either using an independent set of data or Monte Carlo simulations. In the first option data from runs with energies of 8~GeV, 12~GeV, 15~GeV, 25~GeV, 35~GeV, 45~GeV, 80~GeV and 100~GeV have been used. The correction has then been applied to events from independent runs with energies of 10 ~GeV, 20~GeV, 25~GeV, 35~GeV, 45~GeV, 60~GeV and 80~GeV.	The	events from the three runs with energies of 10~GeV, 20~GeV and  60~GeV allow the performance of the correction to be tested also at energies not used to build the looking up table.

The results are shown in figure~\ref{Fig:LeakageCorrection}. The mean energy measured by SiW-ECAL+AHCAL is shifted to the expected beam energy, correcting for the average leakage, with an accuracy better than 1\%, which is comparable to the precision given by the additional information of the TCMT. The RMS90/Mean improvement is up to 25\% at 80~GeV while decreasing towards lower energies together with the impact of the leakage.

\subsection{Conclusion}
The correction was applied to selected pion data and it was tuned either using independent data runs or Monte Carlo simulations. The obtained results are better in case the correction is built from data, due to discrepancies between data and simulations. The correction reduces significantly the impact of the leakage. The mean of the energy distributions after the correction is centered around the expected beam energy with a 1\% or 2\% accuracy, depending on whether the corrections are built from data or from Monte Carlo simulations. Such a precision is approximately the same given by the additional use of the TCMT. The resolution is improved by up to 25\% at high energies, where the impact of the leakage is more pronounced. The improvement decreases towards lower energies together with the impact of the leakage.

% ****************************************************************************
% BIBLIOGRAPHY AREA
% ****************************************************************************

\begin{footnotesize}
% IF YOU DO NOT USE BIBTEX, USE THE FOLLOWING SAMPLE SCHEME FOR THE REFERENCES
% ----------------------------------------------------------------------------

% ----------------------------------------------------------------------------

\end{footnotesize}

% ****************************************************************************
% END OF BIBLIOGRAPHY AREA
% ****************************************************************************


\begin{thebibliography}{99}
% Please replace the numbers for   contribId   and   sessionId
% in the following URL. You can get this information by going to 
% http://indico.cern.ch/confAuthorIndex.py?confId=2628
% and search for your contribution and click on the title
% Be aware: '&amp;' must be replaced by simple '&' as in example below
%------- replace following references ;-)
\bibitem{ILC}J.~Brau, {\it et~al.}, {\it ILC Reference Design Report Volume 1 - Executive Summary}, arXiv:0712.1950 [physics.acc-ph], (2007)
\bibitem{PFA} M. A.~Thomson, {\it Particle flow calorimetry and the PandoraPFA Algorithm}, arXiv:0907.3577 [physics.ins-det], (2009)
\bibitem{AHCAL}C.~Adloff {\it et~al.}, {\it Construction and Commissioning of the CALICE Analog Hadron Calorimeter Prototype}, JINST, vol. 5, p. P05004, 2010.
\bibitem{SiPM} G.~Bondarenko {\it et~al.}, {\it Limited Geiger-mode microcell silicon photodiode: New results}, Nucl. Inst. Meth., vol. A442, pp. 187-192, (2000)
\bibitem{ELECTRON}C.~Adloff {\it et~al.}, {\it Electromagnetic response of a highly granular hadronic calorimeter}, JINST, vol. 6, p. P04003, 2011. arXiv:1012.4343
\bibitem{CAN014} {\it The CALICE Collaboration, Electron data with the CALICE tile AHCAL prototype at the CERN test-beam}, CALICE Analysis Note 014, (2008)
\bibitem{TRACKSEGMENT}S.~Lu, {\it Calibration Studies and the Investigation of Track Segments within Showers with an Imaging Hadronic Calorimeter}, arXiv:0910.3820 [physics.ins-det], (2009)
\bibitem{CAN029}I.~Marchesini {\it Shower leakage in a highly granular calorimeter}, CALICE Analysis Note 029, (2011)
\end{thebibliography}
\end{document}